\documentclass[prd,
               nofootinbib,floatfix,
               fleqn,preprint,12pt,tightenlines]{revtex4} 

\usepackage{amsmath,amssymb,revsymb,graphicx,dcolumn}

\newcommand{\beq}{\begin{equation}}
\newcommand{\eeq}{\end{equation}}
\newcommand{\beqa}{\begin{eqnarray}}
\newcommand{\eeqa}{\end{eqnarray}}
\newcommand{\bsubeqs}{\begin{subequations}}
\newcommand{\esubeqs}{\end{subequations}}

\begin{document}

\begin{widetext}
\noindent
JETP \textbf{125}, 268--277 (2017) \hfill   arXiv:1601.04676
\newline\vspace*{0mm}
\end{widetext}

\title{Relaxation of vacuum energy in $q$-theory
\vspace*{0mm}}

\author{F.R. Klinkhamer}
\email{frans.klinkhamer@kit.edu}
\affiliation{Institute for
Theoretical Physics, Karlsruhe Institute of
Technology (KIT), 76128 Karlsruhe, Germany}

\author{M. Savelainen}
\email{matti.savelainen@helsinki.fi}
\affiliation{ Low Temperature Laboratory, Department of Applied Physics, Aalto University, PO Box 15100, FI-00076 AALTO, Finland}

\author{G.E. Volovik}
\email{volovik@ltl.tkk.fi}
\affiliation{\mbox{Low Temperature Laboratory, Department of Applied Physics,}\\
\mbox{Aalto University, PO Box 15100, FI-00076 AALTO, Finland,}\\
and\\
\mbox{Landau Institute for Theoretical Physics, 
Russian Academy of Sciences,}\\
\mbox{Kosygina 2, 119334 Moscow, Russia}\vspace*{6mm}}

\begin{abstract}
\vspace*{1mm}\noindent
The $q$-theory formalism aims to describe
the thermodynamics and dynamics of the deep quantum vacuum.
The thermodynamics leads to an exact cancellation of the
quantum-field zero-point-energies in equilibrium, which
partly solves the main cosmological constant problem.
But, with reversible dynamics, the spatially-flat
Friedmann-Robertson-Walker universe asymptotically approaches the
Minkowski vacuum only if the Big Bang already started out in an initial equilibrium state. Here, we extend $q$-theory by introducing dissipation
from irreversible processes.
Neglecting the possible instability of a de-Sitter vacuum, we obtain
different scenarios with either a de-Sitter asymptote or collapse
to a final singularity.
The Minkowski asymptote still requires fine-tuning of the
initial conditions. This suggests that,
within the $q$-theory approach,
the decay of the de-Sitter vacuum is a necessary
condition for the dynamical solution of the cosmological constant
problem.
\vspace*{-12mm}
\end{abstract}


\maketitle

\section{Introduction}
\label{sec:Intro}

The dynamics of the quantum vacuum is one of the
major unsolved problems of relativistic quantum field theory (RQFT) and cosmology. The reason is that RQFT and
general relativity (GR) describe processes well below the Planck energy scale, while the deep ultraviolet quantum vacuum at or above the Planck energy scale remains
\textit{terra incognita}. Different regularization schemes to treat the ultraviolet divergences do not help much. Especially troublesome is
the estimate of the vacuum energy density, which represents the so-called cosmological constant problem (CCP)~\cite{Weinberg1988}.
We need an extended theory which allows us to study, in a more general way, the dynamical processes that follow after ``cosmological catastrophes''
have strongly perturbed the deep vacuum. The typical source of such  catastrophes is the vacuum instability caused by different types of gravitational backgrounds (see Refs.
\cite{Polyakov2008,Polyakov2010,KrotovPolyakov2011}
and Ref.~\cite{StarobinskyYokoyama1994} for different results)
or by cosmological phase transitions~\cite{Mukhanov2005}.

Similar processes occur in condensed matter physics, where the
ground state of the system (the ``vacuum'') is, for example, disturbed
by a  quench. After a kick of the system from its equilibrium state, the quantum condensed matter (e.g., quantum liquid or superconductor)
experiences a power-law oscillatory
attenuation of the Bardeen--Cooper--Schrieffer order parameter
\cite{VolkovKogan1974,Barankov2004,Yuzbashyan2005,Yuzbashyan2008,Gurarie2009,Gurarie2015}.
The observation of oscillations emerging in superconductors after a kick has even allowed to resolve the Higgs amplitude mode of the order parameter~\cite{Matsunaga2013,Matsunaga2014}. Different from elementary particle physics, condensed matter physics is able to study
both the infrared macroscopic regime described by the effective
quantum field theory and the corresponding ultra-high-energy regime described by the microscopic physics, the known atomic physics.

The experience with vacuum dynamics in condensed matter physics
has led to a special macroscopic approach called $q$-theory, which incorporates the ultraviolet degrees of freedom of the quantum vacuum into an effective
theory~\cite{KlinkhamerVolovik2008a,KlinkhamerVolovik2008b,%
KlinkhamerVolovik2010,KlinkhamerVolovik2016-JETPL}.
The deep physical vacuum is described in terms of a microscopic dynamic variable -- the vacuum field $q$.
The vacuum variable $q$
is a conserved quantity, which allows for the stabilization of the ground state of the system (the vacuum) in the absence of an environment, i.e,
with zero external pressure. The vacuum in our approach is considered as the Lorentz-invariant analog of the condensed matter system (liquid or solid), which is stable in free space. The variable $q$
is the Lorentz-invariant analog of the particle number density $n$,
whose conservation determines
the thermodynamics and dynamics of many-body systems.

The variable $q$ relevant for the description of the quantum vacuum
occurs in different relativistic theories.
In particular, the vacuum variable $q$ has been formulated
in terms of the 4-form field strength
\cite{DuffNieuwenhuizen1980,Aurilia-etal1980,Hawking1984,HenneauxTeitelboim1984,
Duff1989,DuncanJensen1989,BoussoPolchinski2000,Aurilia-etal2004,Wu2008}.
The difference between our approach~\cite{KlinkhamerVolovik2008a}
and these references is that, instead of taking a quadratic action
term $q^2$, we use a general function $\epsilon(q)$, which allows us to consider an equilibrium vacuum with a non-zero value of $q$.
We shall use this specific realization
of the $q$-field, but the results are not very sensitive to the choice of $q$-variable.
The advantage of $q$-theory is that its equations are universal, in the sense that they essentially do not depend on the origin of the $q$--field and that they naturally modify GR.

Whatever the origin of $q$ may be,
the $q$-theory approach naturally solves the problem of the equilibrium energy. The diverging contribution of the zero-point-energies to the vacuum energy density is automatically
cancelled by the microscopic degrees of freedom.
This cancellation
follows from the Gibbs--Duhem identity, which is applicable to any equilibrium ground state, including the one of the physical vacuum.
As a result, the proper vacuum energy density entering the Einstein gravitational equation is zero in full equilibrium at zero temperature.
Let us consider the illustrative case where the matter sector is represented by a single real scalar field with a non-zero absolute minimum of the potential. Without $q$-field, the vacuum has a large energy density and a corresponding large cosmological constant. However, in equilibrium, the $q$-field automatically compensates this contribution to the vacuum energy density without any fine-tuning~\cite{KlinkhamerVolovik2008a}.
The only assumption is that the quantum vacuum is a
self-sustained system with the energy being an extensive quantity.
Out of equilibrium and/or at nonzero temperature,
the vacuum energy density is determined in the infrared and, hence, is many orders of magnitude smaller than the value suggested by the non-zero absolute minimum of the potential or by the ultraviolet cut-off in RQFT.
The  CCP is then reduced to the problem of the relaxation of the vacuum to its thermodynamic equilibrium.

Up till now, we have considered the classical version of
$q$-theory~\cite{KlinkhamerVolovik2008a},  in which the quantum-dissipative energy exchange between vacuum and matter has been neglected. In the classical theory, the analog of the chemical potential $\mu$
-- the variable thermodynamically conjugate to the variable $q$ --
becomes an integration constant. In a perfect equilibrium vacuum,
$\mu$ has the value $\mu_0$ determined by the microscopic parameters of the physical vacuum, which gives a zero value for the cosmological constant.
After a cosmic catastrophe, the energy density of the perturbed vacuum  may be huge, of the order of the Planck energy scale. Still,  if the cosmic catastrophe occurs in the original Minkowski vacuum (i.e., with
$\mu=\mu_0$), the state with a huge cosmological constant   will relax back to the Minkowski state with zero cosmological
constant (see Figs.~1--5 in Ref.~\cite{KlinkhamerVolovik2008b}).

The drawback of the $q$-theory at the classical level is that,
if the original chemical potential
$\mu$ is unequal to the value $\mu_0$, the vacuum does not relax to the Minkowski vacuum but to a de-Sitter
state(see Fig.~6 in Ref.~\cite{KlinkhamerVolovik2008b}).
This situation is similar
to that of superconductors after a quench
\cite{VolkovKogan1974,Barankov2004,Yuzbashyan2005,Yuzbashyan2008,%
Gurarie2009,Gurarie2015}: if dissipation is neglected, the power-law oscillatory attenuation does not necessarily lead to the equilibrium ground state.

The next step is to extend $q$-theory in order to incorporate quantum-dissipation.
This is why the next step should be to extend $q$-theory in order to incorporate quantum-dissipation and, thus, to allow $\mu$ to relax to its equilibrium value $\mu_0$.
In a full quantum theory, the dynamics of the $q$--field and the accompanying oscillating gravitational field should give rise to particle production and, thus, to the transfer of vacuum energy density to
the energy of the produced matter fields.
At this moment, we cannot discuss the  full quantum theory, but
we can use, instead, a phenomenological extension of $q$-theory, which is  based on the theoretical results of particle production in external fields
\cite{ZeldovichStarobinsky1977,BirrellDavies1982,Kofman1997,DobadoMaroto1999}. The question is if there are conditions under which
the Minkowski vacuum appears as an attractor of the dynamical equations.

Here, we consider the case where the dissipation comes from the time-dependence  of the vacuum field $q(t)$ or from the time-dependence
of the Hubble expansion parameter $H(t)$.
In this way, dissipation is absent not only in the Minkowski vacuum but also in
the de-Sitter vacuum. In other words, we assume that the de-Sitter vacuum is not radiating. (The phenomenological $q$-theory based on the Polyakov mechanism of the radiative instability of the de-Sitter vacuum
\cite{Polyakov2008,Polyakov2010,KrotovPolyakov2011}
has been considered in Ref.~\cite{Klinkhamer2012}.)
Even if the Minkowski and de-Sitter vacua are treated on an equal footing,
it could be that some range of initial conditions prefers the Minkowski vacuum, and we intend to check for this possibility.

It appears that in  $q$-theory there are several equilibrium Minkowski states of the quantum vacuum, which correspond to different equilibrium values
$q_{0}^{(n)}$
of $q$. The present physical vacuum has nonzero
$q_{0}^{(1)}\neq 0$ and a nonzero gravitational coupling constant
$G^{-1}[q_{0}^{(1)}]\neq 0$.
But gravity may be absent or completely modified in the trivial vacuum with $q_{0}^{(2)}=0$,
implying that depending on the microscopic (trans-Planckian) theory of the vacuum
either the gravitational coupling constant vanishes, $G[q_{0}^{(2)}]=0$,
or the inverse gravitational coupling constant vanishes, $G^{-1}[q_{0}^{(2)}]=0$.
For the setup considered, we have found that
the nontrivial Minkowski vacuum with $q_{0}\ne 0$ can only result
from fine-tuning.

Though we found a variety of dynamical behaviors, with or without eternal expansion of the model universe,
we have not found attractor behavior in the investigated regions of parameter space. This negative result suggests that, within the $q$-theory approach,
the decay of the de Sitter vacuum (by radiation or due to instability)
is a necessary condition for the dynamical solution of the CCP.
In a companion paper~\cite{KlinkhamerVolovik2016-MPLA},
we discuss another extension of $q$-theory, which leads to the dynamical preference
of the Minkowski vacuum over the de-Sitter vacuum.

\section{$Q$-theory dynamics}
\label{sec:qtheory}

\subsection{$Q$-theory and the cosmological constant problem}

In the four-form realization~\cite{KlinkhamerVolovik2008a,KlinkhamerVolovik2008b},
the vacuum variable $q$ is represented by the antisymmetric field strength $F_{\kappa\lambda\mu\nu}$ of the three-form gauge field
$A_{\lambda\mu\nu}$~\cite{DuffNieuwenhuizen1980,Aurilia-etal1980}:
\begin{subequations}\label{eq:Fdefinition}
\begin{eqnarray}\label{eq:Fdefinition-q-square}
q^2 &\equiv&- \frac{1}{24}\,
F_{\kappa\lambda\mu\nu}\,F^{\kappa\lambda\mu\nu}\,,
\\[2mm]\label{eq:Fdefinition-covariant}
F_{\kappa\lambda\mu\nu}&\equiv&
\nabla_{[\kappa}A_{\lambda\mu\nu]}=
q\,\sqrt{-g} \,\epsilon_{\kappa\lambda\mu\nu}\,,
\\[2mm]\label{eq:Fdefinition-contravariant}
F^{\kappa\lambda\mu\nu} &=&
q \,\epsilon^{\kappa\lambda\mu\nu}/\sqrt{-g}\,,
\end{eqnarray}
\end{subequations}
where $\nabla_\mu$ is the covariant derivative and a pair of
square brackets around spacetime indices stands for complete anti-symmetrization.
Henceforth, we use natural units with $c=\hbar=k_{B}=1$ and
take the metric signature $(-+++)$.
In \eqref{eq:Fdefinition-covariant},
$\epsilon_{\kappa\lambda\mu\nu}$ is the Levi--Civita symbol
and $q$ is a pseudoscalar.
However, $q$ is not a fundamental pseudoscalar but a composite
pseudoscalar made from the gauge field $A_{\kappa\lambda\mu}$
and the metric $g_{\mu\nu}$
(see also Sec. 2 of Ref.~\cite{KlinkhamerVolovik2016-JETPL} for further discussion).

The action for the vacuum field $q$, the generic matter field $\psi$, and the gravity field $g_{\mu\nu}$ is a generalization
of the one considered in Refs.
 \cite{DuffNieuwenhuizen1980,Aurilia-etal1980,Hawking1984,HenneauxTeitelboim1984,
Duff1989,DuncanJensen1989,BoussoPolchinski2000,Aurilia-etal2004,Wu2008}.
Specifically, a Maxwell-type
term, quadratic in $q$, is replaced by the general function $\epsilon(q)$
and Newton's gravitational constant $G_{N}$
is replaced by a function $G(q)$. The action then reads
\begin{eqnarray}
&&
I= \int_{\mathbb{R}^4}
\,d^4x\, \sqrt{-g}\,\left(\frac{R}{16\pi G(q)} +\epsilon(q)
+\mathcal{L}^{M}(\psi)\right) \,.
\label{eq:actionF}
\end{eqnarray}
Here, we have neglected the possible $q$ dependence of the parameters of the matter Lagrange density $\mathcal{L}^{M}(\psi)$.
The action (\ref{eq:actionF}) contains
two arbitrary functions with the only assumption that  the equilibrium vacuum has a nonzero constant
value $q_{0}$ for the $q$ field, i.e., the vacuum is not ``empty'' ($q=0$).
The cancellation of the vacuum energy density
does not depend on the choice of these two functions.
We have already mentioned
that $q$ in the action \eqref{eq:actionF}
is not a fundamental pseudoscalar field,
as it derives from the gauge field $A_{\mu\nu\rho}$
and the metric field $g_{\mu\nu}$, according to
\eqref{eq:Fdefinition-covariant}.

One important characteristic of $q$-theory is that the microscopic
energy density of the vacuum $\epsilon(q)$, which enters the action,  does not coincide with the thermodynamic vacuum energy density $\rho_{V}(q)$,
which enters the Einstein equation as a cosmological constant. Indeed, the  generalized Einstein equation
is obtained by variation of the action \eqref{eq:actionF} over the metric $g_{\mu\nu}$ and has the following form:
\begin{eqnarray}\label{eq:EinsteinEquationF}
&&
\frac{1}{8\pi G(q)}
\left( R_{\mu\nu}-\frac{1}{2}\,R\,g_{\mu\nu}\right)
\nonumber\\[2mm]&&
+\frac{1}{16\pi}\, q\,\frac{d G^{-1}(q)}{d q}\, {R}\,g_{\mu\nu}
+ \frac{1}{8\pi} \Big( \nabla_\mu\nabla_\nu\,  - g_{\mu\nu}\,
\Box\,\Big)\, G^{-1}(q)
\nonumber\\[2mm]&&
-\left( \epsilon(q) -q\,\frac{d\epsilon(q)}{d q}\right)g_{\mu\nu}
 +T^{M}_{\mu\nu} =0\,,
\end{eqnarray}
where $\Box$ is the invariant d'Alembertian and
$T^{M}_{\mu\nu}$ is the  energy-momentum tensor of the matter
field $\psi$. If the dependence of the gravitational coupling $G$ on $q$ is ignored, this is the standard Einstein equation where the role of the cosmological constant $\Lambda$ is played by the following
vacuum energy density:
\begin{equation}
\Lambda= \rho_{V}(q) = - P_{V}(q) =  \epsilon(q) -q\,\frac{d\epsilon(q)}{d q}   \,.
\label{Lambdaq}
\end{equation}

The vacuum energy density $\rho_{V}(q)$ from \eqref{Lambdaq}
is the analog of the thermodynamic potential
\mbox{$\epsilon - n\,d\epsilon/dn$} in condensed matter physics.
In condensed matter physics, such a thermodynamic potential is nullified in the perfect equilibrium at zero temperature ($T=0$) and in the absence of external forces (i.e., vanishing external pressure, $P=0$).
The nullification follows from the integrated form
of the Gibbs--Duhem relation~\cite{KlinkhamerVolovik2008a},
$\epsilon - n d\epsilon/dn = -P$.
This thermodynamic relation is universal and should also be applicable to a relativistic medium such as the physical vacuum. As a result, any equilibrium state of the physical vacuum at $T=0$ has $\rho_{V}=0$, if it is
assumed that the vacuum belongs to the class of the self-sustained systems,
that is, systems which may exist without external environment (i.e., at $P=0$). This means that the nullification of the cosmological constant in the vacuum is a natural consequence of the thermodynamics of self-sustained systems. Due to the thermodynamic laws, the huge contribution to  $\Lambda$ from the zero-point-energies of the quantum fields~\cite{Weinberg1988}  is automatically cancelled by  the microscopic (trans-Planckian) contribution. This cancellation occurs without fine-tuning.

An example of the microscopic vacuum energy density $\epsilon(q)$ and the
corresponding gravitating vacuum energy density $\rho_{V}(q)$
is given by~\cite{KlinkhamerVolovik2008b}
\begin{subequations}\label{EpsilonExample}
\begin{eqnarray}\label{EpsilonExample-epsilon}
\epsilon(q) &=&\frac{1}{2\chi_{0}}  \left( -\frac{q^2}{q_{0}^2}  +   \frac{1}{3}\;\frac{q^4}{q_{0}^4}\right)\,,
\\[2mm]\label{EpsilonExample-rhoV}
\rho_{V}(q) &\equiv&  \epsilon(q) -q\,\frac{d\epsilon(q)}{d q}
=\frac{1}{2\chi_{0}}\;\frac{q^2}{q_{0}^2}\;  \left( 1 -\frac{q^2}{q_{0}^2}\right)\,,
\end{eqnarray}
\end{subequations}
with nonzero constants $q_{0}$ and $\chi_{0}$.
This example shows two types of equilibrium quantum vacua, both obeying
$ \rho_{V}(q)=0$ in flat spacetime.
One type is the trivial vacuum with $q=0$ and $\mu\equiv d\epsilon/dq=0$. The other type has nonzero equilibrium parameters,
\begin{subequations}\label{EquilibriumVacuum-q0-mu0-chi0}
\begin{eqnarray}
\label{EquilibriumVacuum-q0}
q &=& q_{0}\,,
\\[2mm]
\label{EquilibriumVacuum-mu0}
\mu &=& \mu_0 = -\frac{1}{3\chi_{0}\, q_{0}}
\,,
\\[2mm]
\label{EquilibriumVacuum-chi0}
\chi  &=&  \chi_{0}  \,,
\end{eqnarray}
\end{subequations}
where $\chi \equiv (q^2\,d^2\epsilon/dq^2)^{-1}$
corresponds to the isothermal compressibility~\cite{KlinkhamerVolovik2008a}. The trivial vacuum is an ``empty''   one and gravity may also be absent in this vacuum.
That is why an appropriate example of the $q$-dependence of the gravitational coupling is given by the following function~\cite{KlinkhamerVolovik2008b}:
\begin{equation}
G^{-1}(q)=\frac{1}{G_{N}}\frac{\sqrt{q^2}}{|q_{0}|}  \,,
\label{GExample}
\end{equation}
where $G_{N}$ is Newton's constant in the equilibrium vacuum, i.e., at $q=q_{0}$.
If the trivial vacuum is approached ($q\rightarrow 0$),
we have $G^{-1}(q)\rightarrow 0$ and the Einstein-Hilbert term
in the action \eqref{eq:actionF} vanishes. Here we assume
that $q_0$ provides the cut-off energy scale for the inverse gravitational constant and that $G^{-1}$ vanishes in the trivial vacuum.
Since the effective gravitational constant becomes infinite,  one can expect the formation of curvature singularities in this limit; see
Refs.~\cite{GurovichStarobinsky1979,Starobinsky1981} for details.

In principle, the quantum vacuum can be multi-component and have several nonequivalent nontrivial states. However, in any of these states, the thermodynamics requires nullification of the cosmological constant
in a perfect equilibrium vacuum.

\subsection{Reversible dynamics of vacuum energy and pressure}

In the present article, we are primarily
interested in the dynamical approach of the quantum vacuum to the equilibrium state (or equilibrium states).
In the absence of energy dissipation, the equation for the vacuum variable $q$ is obtained
by the variation of the action \eqref{eq:actionF} over the gauge field $A_{\lambda\mu\nu}$,
\begin{equation}
\nabla_\nu \left[\sqrt{-g} \;\frac{F^{\kappa\lambda\mu\nu}}{q} \left(
\frac{d\epsilon(q)}{d q}+\frac{R}{16\pi} \frac{dG^{-1}(q)}{d q}
\right)   \right]=0\,.
\label{eq:Maxwell}
\end{equation}
In the spatially-flat ($k=0$) Friedmann--Robertson--Walker (FRW) universe,
with the field $q$ depending only on the cosmic time $t$,
the generalized Maxwell Eq.~\eqref{eq:Maxwell} is reduced to
\begin{equation}
\partial_t \left[
\frac{d\epsilon(q)}{d q}+\frac{R}{16\pi} \frac{dG^{-1}(q)}{d q}
\right] =0\,,
\label{eq:Maxwell2}
\end{equation}
which results in an integration constant $\mu$ for the solution,
\begin{equation}
\frac{d\epsilon(q)}{d q}+\frac{R}{16\pi} \frac{dG^{-1}(q)}{d q}
=\mu \,.
\label{eq:Maxwell3}
\end{equation}
In condensed matter physics, the integration constant $\mu$ corresponds to the fixed chemical potential.

The fate of the expanding universe after a cosmic catastrophe depends on the  value of this integration constant $\mu$~\cite{KlinkhamerVolovik2008b}.
The universe relaxes to the equilibrium Minkowski vacuum only if
$\mu=\mu_0$ with $\mu_0$ as given by \eqref{EquilibriumVacuum-mu0}
for the \textit{Ansatz} \eqref{EpsilonExample-epsilon}.
For $\mu\neq \mu_0$, the solutions of the dynamic equations have a de-Sitter asymptote, with a Hubble constant $H$ determined by $\mu$. Hence, if reversible dynamics of the vacuum is used, the cosmological constant problem is replaced by another problem~\cite{KlinkhamerVolovik2010}:
why does $\mu$ have the ``correct'' value?

This is the reason for introducing dissipation into the equation for the vacuum variable $q$. Then, the effective chemical potential
$\mu_\text{eff}$ is no
longer an integration constant and may relax.

\subsection{Irreversible dynamics of vacuum energy and pressure}
\label{subsec:Dissipation function}

Phenomenologically, we introduce dissipation by adding a source term to
\eqref{eq:Maxwell2}:
\begin{equation}
\partial_t\left[\frac{d\epsilon(q)}{d q}+\frac{R}{16\pi} \frac{dG^{-1}(q)}{d q}\right]
=S \,.
\label{eq:Maxwell4}
\end{equation}
The role of the function $S$ can be best understood if, for the moment, we neglect the dependence of the gravitational coupling $G$ on $q$. Then, from the
Friedmann equation
[based on the Einstein equation \eqref{eq:EinsteinEquationF}]
and Eq.~\eqref{eq:Maxwell4} for the vacuum, we obtain the following
evolution equations for the energy densities of vacuum and matter:
\begin{subequations}\label{eq:exchange}
\begin{eqnarray}
\partial_t   \rho_{V} &=&  -q\, S\,,
\\[2mm]
\partial_t  \rho_{M} &=&-3H\, \Big(P_{M}+\rho_{M}\Big) +q\, S\,.
\end{eqnarray}
\end{subequations}
This demonstrates that $q\, S$ describes the dissipation of vacuum energy
density into matter excitations.
The energy exchange between vacuum and matter is caused by particle production, which takes place due to either a time-dependent gravitational
environment~\cite{ZeldovichStarobinsky1977,BirrellDavies1982,%
DobadoMaroto1999} or a parametric resonance caused by oscillations of the fields~\cite{Kofman1997,Felice2012}.

Here, we follow the lead of condensed matter physics, where the dissipation function can be expressed in terms of a quadratic form of the time derivatives. In our case, we have
\begin{equation}
 q\, S = \Gamma_q  (\partial_t   q)^2 +  \Gamma_H  (\partial_t   H)^2  \,,
\label{S}
\end{equation}
with nonnegative constants $\Gamma_q$ and $\Gamma_H$.
The second term on the right-hand side of
\eqref{S} has a parallel with the $R^2$ term for particle production in Ref.~\cite{ZeldovichStarobinsky1977}, where  $R$ is the Ricci scalar.  For the spatially-flat FRW universe,
we have $R^2=36\,(\partial_t H+2H^2)^2$,
which differs from $36\,(\partial_t H)^2$ by a total time derivative:
$ a^3\,[R^2-36(\partial_t H)^2]=  \partial_t [48\,(\partial_t a)^3]$. The second term on the right-hand side of \eqref{S} can be
written in terms of the Riemann tensor as the combination of $R^2$ and the Gauss-Bonnet term
$E=R^2-4 R^{\mu\nu} R_{\mu\nu}+ R^{\mu\nu\rho\sigma} R_{\mu\nu\rho\sigma}$, the latter also giving a total time derivative
as $E_\text{flat-FRW}= 24\,H^2\,(\partial_t H+H^2)$.
Specifically, we have
\begin{equation}
\frac{1}{36}\,\Big(R^2 - 6E\Big)\,\Big|_\text{flat-FRW} =  (\partial_t H)^2\,.
\label{eq:dotHsquare}
\end{equation}

The crucial property of the source term \eqref{S} is that it does not discriminate between Minkowski and de-Sitter vacua: the dissipation vanishes for both vacua. The particular phenomenological term \eqref{S}
can only be relevant if the de-Sitter vacuum is stable.
The issue of the stability of the de-Sitter vacuum remains an unsolved problem. The possible instability of the de-Sitter vacuum is discussed in Refs.~\cite{Polyakov2008,Polyakov2010,KrotovPolyakov2011,Akhmedov2014}.
An alternative point of view on the fate
of the de-Sitter vacuum is given by Ref.~\cite{StarobinskyYokoyama1994}.
Here, we intend to study whether or not the dissipative
source term \eqref{S} dynamically leads to the Minkowski asymptote.
In a companion paper~\cite{KlinkhamerVolovik2016-MPLA}, we have
considered the source term $S\propto |H|\,R^2$,
which does discriminate between Minkowski and de-Sitter vacua.

\section{Model equations}
\label{sec:Model-equations}

As in Ref.~\cite{KlinkhamerVolovik2008b}, we use dimensionless variables
$f$,   $k$, $h$, and $\tau$,
\begin{subequations}\label{eq:Dimensionless1}
\begin{eqnarray}
q        &\equiv& f \, q_{0}\,,
\\[2mm]
G^{-1}(q)&\equiv& k(f) \, |q_{0}|\,,
\\[2mm]
H &\equiv& h/\sqrt{\chi_{0} \, |q_{0}|}\,,
\\[2mm]
t &\equiv&  \tau \,\sqrt{\chi_{0}\, |q_{0}|}\,,
\end{eqnarray}
\end{subequations}
with the following \textit{Ansatz} based on \eqref{GExample}:
\begin{eqnarray}
\label{eq:k-Ansatz}
k(f) &=& \frac{8\pi}{3}\,|f| = \frac{8\pi}{3}\,f \,,
\end{eqnarray}
if we assume that $f$ stays positive.
Moreover, a dot and a prime will stand for a derivative
with respect to $\tau$ and $f$, respectively.

In terms of these dimensionless variables,
the vacuum, Friedmann, and matter equations have the following form:
\begin{subequations}\label{eq:3eqsFRWdim3}
\begin{eqnarray}
\ddot{h} + 4 \;h \;\dot{h}&=&
\dot{f} \;\epsilon^{\prime\prime}-\widetilde{s} \,,
\label{eq:MaxwellFRWdim3}
\\[2mm]
h \;\dot{f} + f\; h^2 &=& \widetilde{r}_{V}+r_{M}\,,
\label{eq:EinsteinFRWdim3}\\[2mm]
\dot{r}_{M}+3\,h\,\big(1+w_{M}\big)\,r_{M}&=&f\;\widetilde{s} \,,
\label{eq:MatterFRWdim3}
\end{eqnarray}
\end{subequations}
with the effective vacuum energy density
\begin{subequations}\label{eq:3eqsFRWdim3-functions}
\begin{eqnarray}
\label{eq:3eqsFRWdim3-widetilde-rV}
\widetilde{r}_{V}  &=&
\epsilon -f\,\epsilon'+ f\; (\dot{h}+2 h^2)
\end{eqnarray}
and the \textit{Ansatz} functions
\begin{eqnarray}
\label{eq:3eqsFRWdim3-widetilde-s}
\widetilde{s} &=&
\gamma_{f}\; (\dot{f})^2 + \gamma_{h}\; (\dot{h})^2 \,,
\\[2mm]
\label{eq:3eqsFRWdim3-widetilde-epsilon}
\epsilon &=& \frac{1}{2}\; \left(-f^2 +\frac{1}{3}\,f^4\right)\,,
\end{eqnarray}
\end{subequations}
where \eqref{eq:3eqsFRWdim3-widetilde-s} and
\eqref{eq:3eqsFRWdim3-widetilde-epsilon} correspond
to the previous \textit{Ans\"{a}tze} \eqref{S} and
\eqref{EpsilonExample-epsilon}, respectively.
The ODEs \eqref{eq:3eqsFRWdim3} can also be obtained from those of
Sec.~IV in Ref.~\cite{KlinkhamerVolovik2008b}
by using Eq.~(4.1) of that reference,
which corresponds to Eq.~\eqref{eq:Maxwell3} here,
to eliminate $\mu$.

Now, define the dimensionless effective chemical potential
(in dimensional units  $\mu_\text{eff}\equiv u_\text{eff} /\chi_{0}q_{0}$),
\begin{equation}
u_\text{eff} \equiv \epsilon'-\dot{h}-2 h^2\,,
\label{eq:u-eff}
\end{equation}
and obtain
\begin{equation}
\widetilde{r}_{V}=\epsilon -u_\text{eff}\,f\,,
\label{eq:rVtilde-with-u-eff}
\end{equation}
which resembles the flat-spacetime result without
dissipation~\cite{KlinkhamerVolovik2008a}, $r_{V}(f)=\epsilon(f) -u\,f$
[or, using dimensional variables, $\rho_{V}(q)=\epsilon(q) -\mu\,q$].
With $u_\text{eff}$ from \eqref{eq:u-eff},
the first ODE \eqref{eq:MaxwellFRWdim3} can be read as
\begin{equation}
\frac{du_\text{eff}}{d\tau}=\widetilde{s}\,,
\label{eq:u-eff-ODE}
\end{equation}
which allows for the relaxation of the vacuum energy density to
the equilibrium value.
Indeed, the ODE \eqref{eq:MaxwellFRWdim3} can also be written as
\begin{equation}
\frac{d\widetilde{r}_{V}}{d\tau}=
-f\,\widetilde{s} + \dot{f}\,(\dot{h}+2 h^2)\,.
\label{eq:rV-tilde-ODE}
\end{equation}

The ODEs~\eqref{eq:3eqsFRWdim3} have an exact solution with
constant functions $f$, $h$,  and $r_{M}$:
\begin{subequations}\label{eq:3eqsFRWdim3-dSsol}
\begin{eqnarray}
f(\tau)&=& \overline{f}=\text{const} \,,
\label{eq:3eqsFRWdim3-dSsol-f}
\\[2mm]
\Big(h(\tau)\Big)^2  &=&
\left[\frac{d\epsilon}{df}  - \frac{\epsilon}{f} \right]_{f=\overline{f}}\,,
\label{eq:3eqsFRWdim3-dSsol-h2}
\\[2mm]
r_{M}(\tau)&=&0\,.
\label{eq:3eqsFRWdim3-dSsol-rM}
\end{eqnarray}
\end{subequations}
This de-Sitter solution does not exist for values of
$\overline{f}$ which make the right-hand side of \eqref{eq:3eqsFRWdim3-dSsol-h2} negative.
In that case, there may exist another
type of asymptotic solution, as will be seen in the next section.

\section{Numerical results and discussion}
\label{sec:Numerical-results}

We have numerically investigated the model-universe dynamics
from the ODEs \eqref{eq:3eqsFRWdim3}, either
with or without dissipation.
Without dissipation, the universe approaches, in general, a de-Sitter vacuum  with a Hubble constant $H$ determined by the original chemical potential
$\mu$ which plays the role of an integration constant.
The Minkowski vacuum as the final state of the evolution is obtained
for a fine-tuned value of the chemical potential, $\mu=\mu_0$.
The Minkowski vacuum is approached, in particular, if the cosmic catastrophe occurs already in the original equilibrium state or if the catastrophe is a local event which does not disturb the chemical potential at spatial infinity,  $\lim_{|\vec{x}|\rightarrow \infty}\mu(\vec{x})=\mu_0$.

\begin{figure*}[p]
\vspace*{0mm}
\begin{center}
\includegraphics[width=0.60\textwidth]{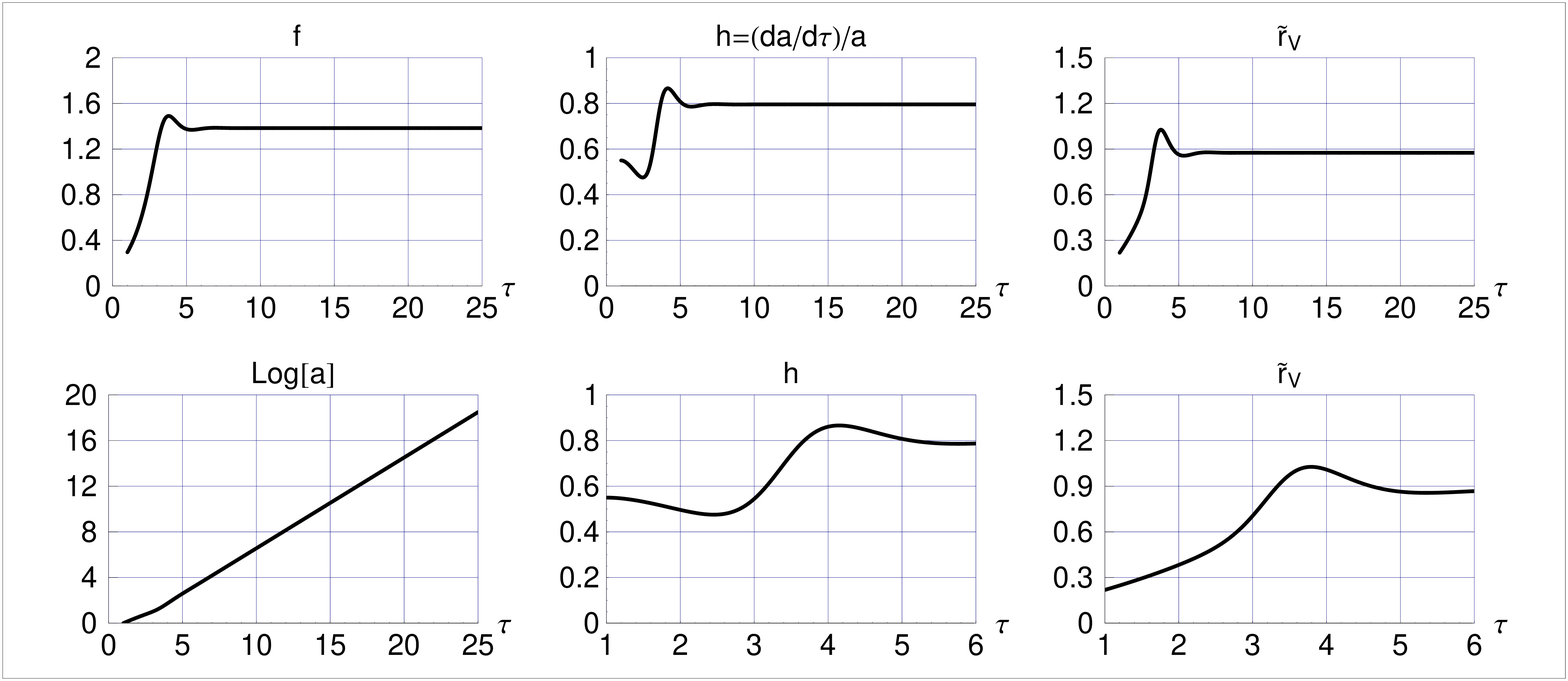}
\end{center}
\vspace*{-5mm}
\caption{Numerical solution for the evolution of the universe
after a cosmic catastrophe, neglecting dissipation. The model parameters
of the ODEs \eqref{eq:3eqsFRWdim3}
with auxiliary functions \eqref{eq:3eqsFRWdim3-functions}
are $(w_{M},\,\gamma_{f},\,\gamma_{h})=(1/3,\, 0, \,0)$ and the
boundary conditions at $\tau=1$  are
$\{a(1),\, h(1),\, \dot{h}(1),\, f(1),\, r_{M}(1)\}=
\{1,\, 0.55,\,0,\, 0.2953,\,0\}$. With these parameters and
boundary conditions, the matter energy density vanishes exactly,
$r_{M}(\tau)=0$ for $\tau \geq 1$.
}
\label{fig:no-dissipation}  
\vspace*{0mm}
\begin{center}
\includegraphics[width=0.60\textwidth]{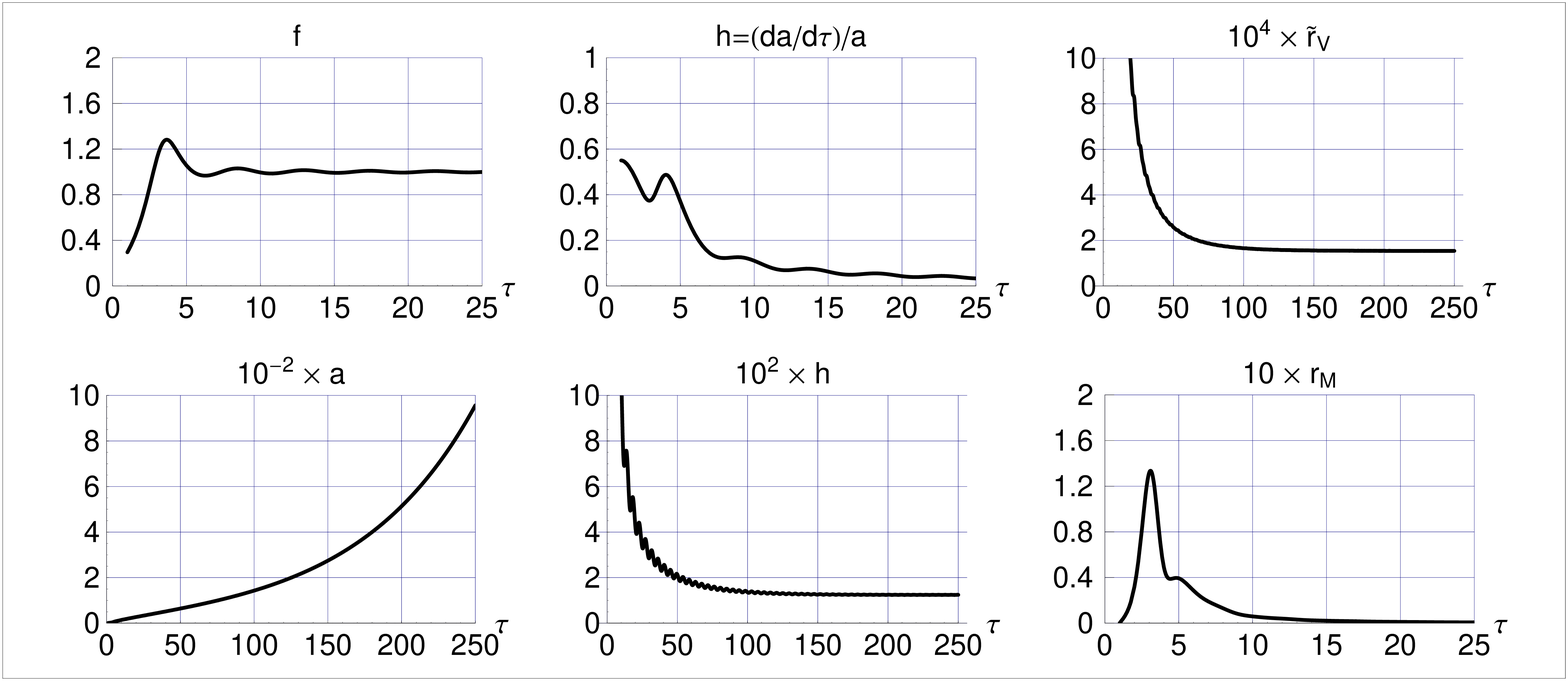}
\end{center}
\vspace*{-5mm}
\caption{Numerical solution for the evolution of the universe after a
cosmic catastrophe, taking into account dissipation, which leads to energy exchange between vacuum and matter. The model parameters
of the ODEs \eqref{eq:3eqsFRWdim3}
are $(w_{M},\,\gamma_{f},\,\gamma_{h})=(1/3,\, 1, \,1)$.
The
boundary conditions at $\tau=1$  are the same as in
Fig.~\ref{fig:no-dissipation},
$\{a(1),\, h(1),\, \dot{h}(1),\, f(1),\, r_{M}(1)\}=
\{1,\, 0.55,\,0,\, 0.2953,\,0\}$.
}
\label{fig:dissipation-asymp-deSitter}  
\vspace*{0mm}
\begin{center}
\includegraphics[width=0.60\textwidth]{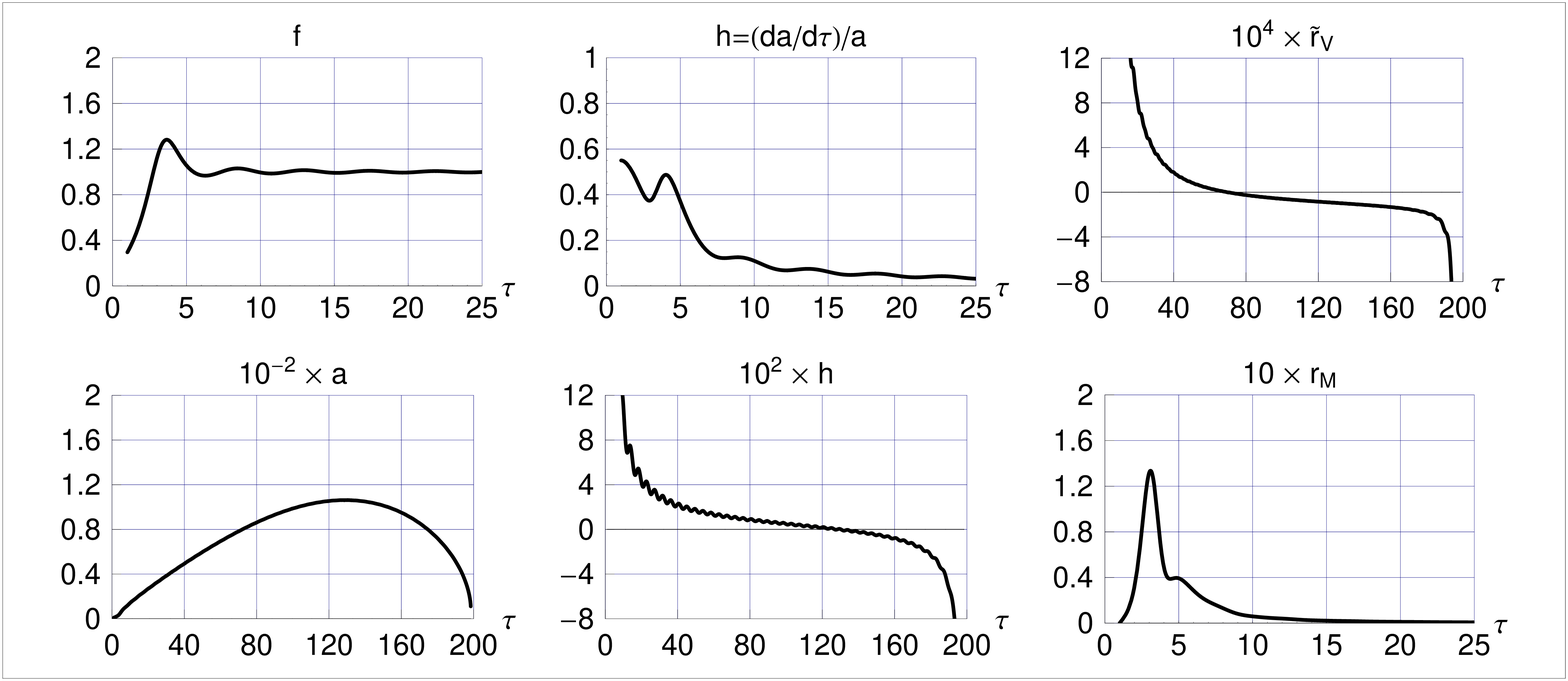}
\end{center}
\vspace*{-5mm}
\caption{Numerical solution of the ODEs \eqref{eq:3eqsFRWdim3},
taking into account dissipation. The model parameters are
the same as in
Fig.~\ref{fig:dissipation-asymp-deSitter},
$(w_{M},\,\gamma_{f},\,\gamma_{h})=(1/3,\, 1, \,1)$, but
the boundary conditions at $\tau=1$  are different,
$\{a(1),\, h(1),\, \dot{h}(1),\, f(1),\, r_{M}(1)\}=
\{1,\, 0.55,\,0,\, 0.2952,\,0\}$. The scale factor $a(\tau)$
of the numerical solution drops to a value
$\text{O}(10)$ at $\tau\approx 198$.
}
\label{fig:dissipation-recollapse}  
\end{figure*}

Figures~\ref{fig:no-dissipation}
and \ref{fig:dissipation-asymp-deSitter}
demonstrate the evolution starting with the same initial conditions
but without and with dissipation, respectively.
Without dissipation
(Fig.~\ref{fig:no-dissipation}),
the universe approaches the de-Sitter vacuum with a Hubble constant
$H$ determined by the integration constant. The effective chemical potential $u_\text{eff}$ starts and keeps
the value $-0.883133$, which is far away from the Minkowski
equilibrium value $-1/3$  (cf. Ref.~\cite{KlinkhamerVolovik2008b}).

With dissipation  and certain initial conditions
(Fig.~\ref{fig:dissipation-asymp-deSitter}),
we still end up with a de-Sitter universe.
The effective chemical potential $u_\text{eff}$ changes
from an initial
value $u_\text{eff}(1)=-0.883133$  to a value $u_\text{eff}(100)=-0.333499$,
close to the Minkowski equilibrium value $-1/3$.
Changing the initial conditions somewhat, we get in
Fig.~\ref{fig:dissipation-recollapse}
a similar
drop of the effective chemical potential $u_\text{eff}$ from a value $u_\text{eff}(1)=-0.883133$  to a value $u_\text{eff}(98)=-0.333276$.
But the further evolution of the model universe is very different:
the expansion continues forever
(Fig.~\ref{fig:dissipation-asymp-deSitter})
or the expansion stops and the universe starts to
contract
(Fig.~\ref{fig:dissipation-recollapse}).

In the four-dimensional space of initial conditions,
$M_\text{in}$ $=$ $\{h(1),\,\dot{h}(1),\,f(1),\,r_{M}(1)\}$,
there is indeed a \emph{separatrix}
which divides regions with or without asymptotic-de-Sitter behavior.
For the source term \eqref{eq:3eqsFRWdim3-widetilde-s}
with $\gamma_f=\gamma_h=1$, we have first considered
the slice $\{\dot{h}(1),\,r_{M}(1)\}=\{0,\,0\}$ and
have found a separatrix line between expansion
and re-collapse behavior in the plane spanned by $h(1)$ and $f(1)$ values (see Table~\ref{tab:Separatrix}).
For slightly different slices of $\{\dot{h}(1),\,r_{M}(1)\}$, the separatrix line moves little
(see caption of Table~\ref{tab:Separatrix}),
and we conclude that the separatrix
is a three-dimensional submanifold of $M_\text{in}$, at least, for the ranges considered.
It remains to be seen if the separatrix extends over the
whole of the four-dimensional space $M_\text{in}$.
In addition,  a detailed study is needed of
the final singularity where it occurs
(see the Appendix for some preliminary remarks).

\begin{table*}[t] 
\vspace*{0cm}
\begin{center}
\caption{Depending on the boundary conditions, the
ODEs \eqref{eq:3eqsFRWdim3} may or may not give an asymptotic de-Sitter spacetime. The model parameters are
$(w_{M},\,\gamma_{f},\,\gamma_{h})=(1/3,\, 1, \,1)$.
For the slice $\{\dot{h}(1),\,r_{M}(1)\}=\{0,\, 0\}$,
the results are shown for $h(1)$ values ranging between $0.525$ and $0.7$
and  $f(1)$ values ranging between $0.24$ and $0.49$:
Y/N stands for a Yes/No answer to the question if there occurs
an  asymptotic de-Sitter spacetime.
The same Y/N pattern is obtained for a slice with
$\{\dot{h}(1),r_{M}(1)\}=\{-1/100,\, 0\}$ and for a slice with
$\{\dot{h}(1),r_{M}(1)\}=\{0,\, 1/50\}$.
This suggest that the separatrix
is three-dimensional, at least, for the ranges considered.
\vspace*{5mm}}
\label{tab:Separatrix}
\renewcommand{\tabcolsep}{1.1pc}    
\renewcommand{\arraystretch}{1.1}   
\begin{tabular}{c||c|c|c|c|c|c}
       &\multicolumn{6}{c}{$h(1)$}\\
$f(1)$ & $0.5250$ & $0.5375$ & $0.55$ & $0.60$ & $0.65$ & $0.70$\\
\hline
\hline
$0.24$ & N & N & N & Y & Y & Y \\
\hline
$0.29$ & N & N & N & Y & Y & Y \\
\hline
$0.34$ & N & Y & Y & Y & Y & Y \\
\hline
$0.39$ & Y & Y & Y & Y & Y & Y \\
\hline
$0.44$ & Y & Y & Y & Y & Y & Y \\
\hline
$0.49$ & Y & Y & Y & Y & Y & Y \\
\hline
\end{tabular}
\end{center}
\vspace*{0mm}
\vspace*{5mm}
\vspace*{0mm}
\begin{center}
\caption{Numerical solutions of the
ODEs \eqref{eq:3eqsFRWdim3} with boundary conditions
close to the separatrix of Table~\ref{tab:Separatrix}.
The model parameters are
$(w_{M},\,\gamma_{f},\,\gamma_{h})=(1/3,\, 1, \,1)$.
For the slice $\{h(1),\,\dot{h}(1),\,r_{M}(1)\}=\{0.525,\,0,\, 0\}$,
the initial value of $f(1)$ is fine-tuned towards obtaining the
Minkowski vacuum
with $h(\infty)=\widetilde{r}_{V}(\infty)=r_{M}(\infty)=0$.
Similar results have been obtained for
the slice $\{h(1),\,\dot{h}(1),\,r_{M}(1)\}=\{0.55,\,0,\, 0\}$,
see also
Figs.~\ref{fig:dissipation-asymp-deSitter} and
\ref{fig:dissipation-recollapse}.
\vspace*{5mm}}
\label{tab:fine-tuning}
\renewcommand{\tabcolsep}{1.1pc}    
\renewcommand{\arraystretch}{1.1}   
\begin{tabular}{c||cccc}
$f(1)$ & $h(200)$ & $f(200)$ & $\widetilde{r}_{V}(200)$ & $r_{M}(200)$ \\
\hline
\hline
$0.352$ & $-$ & $-$ & $-$ & $-$ \\
\hline
$0.353$ & $0.0362$ & $1.00131$ & $0.00132$ & $5 \times 10^{-12}$ \\
\hline
$0.354$ & $0.0510$ & $1.00259$ & $0.00260$ & $3 \times 10^{-14}$ \\
\hline
$0.355$ & $0.0615$ & $1.00376$ & $0.00379$ & $5 \times 10^{-15}$ \\
\hline
$0.36$ & $0.0952$ & $1.00895$ & $0.00915$ & $0$ \\
\hline
$0.37$ & $0.136$ & $1.0180$ & $0.0188$ & $0$ \\
\hline
$0.38$ & $0.164$ & $1.0260$ & $0.0277$ & $0$ \\
\hline
$0.39$ & $0.187$ & $1.0334$ & $0.0363$ & $0$ \\
\hline
\end{tabular}
\end{center}
\vspace*{0mm}
\end{table*}

The behavior of the solutions close to the separatrix has
already been shown in
Figs.~\ref{fig:dissipation-asymp-deSitter}
and \ref{fig:dissipation-recollapse},
which have slightly different
values of the boundary condition $f(1)$.
Approaching the separatrix from the de-Sitter side, the
asymptotic vacuum energy density approaches zero from above
and the asymptotic-de-Sitter universe can get arbitrarily close
to the Minkowski universe with vanishing vacuum energy density.
In this way, the Minkowski vacuum is obtained by fine-tuning
of the initial conditions.
Still, the Minkowski vacuum does not originate from a
point in the  four-dimensional space of initial conditions
$M_\text{in}$ but rather from a three-dimensional
submanifold and the fine-tuning is only one-dimensional
(if started close enough to the separatrix and on the
de-Sitter side; see  Table~\ref{tab:fine-tuning}).

\section{Conclusion}
\label{sec:Conclusion}

In this article,
we have introduced dissipation into $q$-theory, in order to study the irreversible relaxation of the vacuum energy density. We have used a general phenomenological approach in terms of a dissipation function, describing the decay of the coherent motion of the quantum vacuum into the incoherent degrees of freedom of the matter fields. We have chosen a dissipation function which does not discriminate between Minkowski and de-Sitter vacua, as the dissipation function vanishes for both vacua. This approach assumes that the de-Sitter universe is not radiating.

Specifically, we have used a simple model of the deep quantum vacuum,  described by a single dynamic variable $q$ expressed in terms of the four-form field strength $F$. Even with this simplification, $q$-theory demonstrates different scenarios for the behavior of the universe depending on the parameters of the system and the initial conditions.
This includes the relaxation to a de-Sitter vacuum or the
collapse to a final singularity. The last type of behavior may correspond to
a scenario where the universe cycles through a finite or infinite number of Big Bangs, each followed by expansion and contraction. In our case, different from the scenario discussed in Ref.~\cite{SteinhardtTurok2002}, the vacuum energy
density would decrease after each cycle due to dissipation.

We have not found a Minkowski attractor, but it is still possible that
a Minkowski attractor exists in some region of parameter space for
the case of multi-component $q$-fields. If that does not happen,
then the fact that the present universe is very
close to equilibrium can be explained, within the $q$-theory approach,
only if we accept that the de-Sitter vacuum is radiating.
For the case of a nonzero dissipation function in a de-Sitter universe,
the relaxation to the Minkowski vacuum has been considered in the
companion paper~\cite{KlinkhamerVolovik2016-MPLA}.

\section*{\hspace*{-5mm}ACKNOWLEDGMENTS}
\noindent
The work of MS has been supported by the Academy of Finland
(Project No. 284594).
The work of  GEV has been supported by the European Research Council
(ERC) under the European Union's Horizon 2020 research and innovation programme (Grant Agreement No. 694248).

\begin{appendix}
\section{}
\label{app:Final-singularity}

In this appendix, we take a closer look at the final singularity found numerically in Sec.~\ref{sec:Numerical-results}.
If we make the approximation that $f(\tau)$
is constant, we are able to obtain an analytic
solution close to the singularity.

Equation~\eqref{eq:MaxwellFRWdim3} then gives
\begin{subequations}\label{eq:sing-ODEs}
\begin{eqnarray}\label{eq:sing-ODEs-fbar}
f &=& \overline{f} = \text{const}\,,
\\[2mm]\label{eq:sing-ODEs-hddot}
\ddot{h} + 4 \;h \;\dot{h}&=& -\gamma_{h}\; (\dot{h})^2 \,.
\end{eqnarray}
Next, take the time derivative of \eqref{eq:EinsteinFRWdim3}
for $f = \overline{f}$
and use Eqs.~\eqref{eq:sing-ODEs-hddot} and \eqref{eq:MatterFRWdim3},
with the result
\begin{eqnarray}\label{eq:sing-ODEs-rM}
r_{M}&=& -\frac{2}{3}\;\frac{1}{1+w_{M}}\; \overline{f}\;\dot{h} \,.
\end{eqnarray}
\end{subequations}
Hence, we only need to solve the single ODE~\eqref{eq:sing-ODEs-hddot}
for $h(\tau)$, which
then gives $r_{M}(\tau)$ from \eqref{eq:sing-ODEs-rM}.

The second-order ODE~\eqref{eq:sing-ODEs-hddot} requires
two boundary conditions, for example, two initial conditions
on $h(\tau_\text{in})$ and $\dot{h}(\tau_\text{in})$.
These two initial conditions are, however, not independent,
but are related to the value of $\overline{f}$ by
\begin{eqnarray}\label{eq:sing-ODEs-bc-constraint}
\frac{1/3+w_{M}}{1+w_{M}}\;\overline{f}\;\dot{h}(\tau_\text{in})
+\overline{f}\;\Big(h(\tau_\text{in})\Big)^2&=&
\Big[ f\,\epsilon' - \epsilon \Big]_{f=\overline{f}} \,.
\nonumber\\[2mm]&&
\end{eqnarray}
This constraint equation corresponds to the original ODE \eqref{eq:EinsteinFRWdim3} for $f = \overline{f}$ and
evaluated at $\tau=\tau_\text{in}$, where $r_{M}(\tau_\text{in})$
has been eliminated by use of \eqref{eq:sing-ODEs-rM}.

In the following, we set $\gamma_{h}=1$ for simplicity.
In \eqref{eq:sing-ODEs-hddot}, we first neglect
the $4 \,h \,\dot{h}$ term on the left-hand side,
which will be justified \mbox{\textit{a posteriori}.}
Then, the resulting ODE for $\dot{h}(\tau)$ is easily solved
and one further integration gives the solution for $h(\tau)$.
For the time interval $\tau < \tau_\text{sing}$, this $h(\tau)$ solution
and the corresponding $r_{M}(\tau)$ solution are given by
\begin{subequations}\label{eq:sing-ODEs-hsol-rMsol}
\begin{eqnarray}\label{eq:sing-ODEs-hsol}
h(\tau)&=&  c_{h} + \ln(\tau_\text{sing} -\tau)\,,
\\[2mm]\label{eq:sing-ODEs-rMsol}
r_{M}(\tau)&=& \frac{2}{3}\;\frac{1}{1+w_{M}}\;\overline{f}\;
\frac{1}{\tau_\text{sing} -\tau} \,,
\end{eqnarray}
\end{subequations}
with constants $c_{h}$ and $\tau_\text{sing}$.
There is a similar solution for $\tau > \tau_\text{sing}$.
With \eqref{eq:sing-ODEs-hsol},
it can now be checked that the $\ddot{h}$ term on the
left-hand side of \eqref{eq:sing-ODEs-hddot}
dominates over the $4\,h\,\dot{h}$ term, namely, the first term
goes as
$(\tau_\text{sing}-\tau)^{-2}$ and the second
as $(\tau_\text{sing} -\tau)^{-1}\,\ln(\tau_\text{sing} -\tau)$.
The scale parameter $a(\tau)$, defined by
$\dot{a}/a=h$, is readily obtained
from \eqref{eq:sing-ODEs-hsol},
\begin{eqnarray}\label{eq:sing-ODEs-asol}
a(\tau)&=& C\;\exp\Big[ (c_h-1)\,\tau
- (\tau_\text{sing} -\tau) \ln(\tau_\text{sing} -\tau)\Big]\,,
\nonumber\\[2mm]&&
\end{eqnarray}
for $\tau < \tau_\text{sing}$ and with an arbitrary constant $C>0$.
This scale parameter $a(\tau)$ does not vanish
as $\tau$ approaches $\tau_\text{sing}$ from below.
Still, there is a physical singularity
as $\tau \uparrow \tau_\text{sing}$ with, for example,
both the Ricci scalar $R$ and the matter energy density $r_{M}$
diverging as $(\tau_\text{sing} -\tau)^{-1}$.

Incidentally, we also have another type of solution
of the ODEs~\eqref{eq:sing-ODEs}, namely, the
de-Sitter solution:
\begin{subequations}\label{eq:sing-ODEs-dSsol}
\begin{eqnarray}
f &=& \overline{f} = \text{const}\,,
\\[2mm]
h &=& \overline{h} = \text{const}\,,
\\[2mm]
\dot{r}_{M}&=& 0 \,,
\end{eqnarray}
\end{subequations}
as discussed at the end of Sec.~\ref{sec:Model-equations}.
The two types of solutions are distinguished by the
$\dot{h}$ boundary condition for appropriate values of $\overline{f}$:
$\dot{h}(\tau_\text{in})=0$
gives the de-Sitter solution \eqref{eq:sing-ODEs-dSsol}
for $\tau > \tau_\text{in}$
and $\dot{h}(\tau_\text{in})<0$ gives the
asymptotic singular solution \eqref{eq:sing-ODEs-hsol-rMsol}
for $\tau_\text{in} < \tau< \tau_\text{sing}$.

Expanding on the last statement,
observe that the constraint \eqref{eq:sing-ODEs-bc-constraint}
allows for a heuristic understanding
of the two different types of solutions found numerically
in Sec.~\ref{sec:Numerical-results}.
If the asymptotic value $\overline{f}$ is such that the
right-hand side of \eqref{eq:sing-ODEs-bc-constraint} is positive,
then a de-Sitter vacuum is possible.
But, if the asymptotic value $\overline{f}$ is such that the
right-hand side of \eqref{eq:sing-ODEs-bc-constraint} is negative,
then a de-Sitter vacuum with $\dot{h}=0$ is strictly impossible and
we are led to the asymptotic singular solution.

In closing, we have two remarks on the relevance of
the asymptotic singular solution as given by
\eqref{eq:sing-ODEs-hsol-rMsol}.
First, this singular analytic solution gives only a qualitative
description of the numerical solution, because
$f_\text{num}(\tau)$ appears to be not perfectly constant.
Second, the physical relevance of the singular solution
relies on the applicability of the \textit{Ansatz}
\eqref{S} for the dissipation function, which may not be
valid for the conditions at the final singularity.

\end{appendix}



\begin{thebibliography}{99}

\bibitem{Weinberg1988}
S. Weinberg,
``The cosmological constant problem,''
Rev. Mod. Phys.  {\bf 61}, 1 (1989).

\bibitem{Polyakov2008}
A.M. Polyakov,
``De Sitter space and eternity,''
Nucl. Phys. B \textbf{797}, 199 (2008), arXiv:0709.2899.


 \bibitem{Polyakov2010}
A.M. Polyakov,
``Decay of vacuum energy,''
Nucl. Phys. B \textbf{834}, 316 (2010),
arXiv:0912.5503.

\bibitem{KrotovPolyakov2011}
D.~Krotov and A.M.~Polyakov,
``Infrared Sensitivity of unstable vacua,''
Nucl.\ Phys.\ B {\bf 849}, 410 (2011),
arXiv:1012.2107.  


\bibitem{StarobinskyYokoyama1994}
A.A. Starobinsky and J. Yokoyama,
``Equilibrium state of a selfinteracting scalar field
  in the de Sitter background,''
Phys.\ Rev.\ D {\bf 50}, 6357 (1994),
arXiv:astro-ph/9407016.

\bibitem{Mukhanov2005}
V. Mukhanov,
\emph{Physical Foundations of Cosmology}
(Cambridge University Press, Cambridge, England, 2005).


\bibitem{VolkovKogan1974}
A.F. Volkov and  S.M. Kogan,
``Collisionless relaxation of the energy gap in superconductors,''
Sov. Phys. JETP {\bf 38}, 1018 (1974).

\bibitem{Barankov2004}
R.A. Barankov, L.S. Levitov, and B.Z. Spivak,
``Collective Rabi oscillations and solitons in a time-dependent BCS pairing problem,''
Phys. Rev. Lett. {\bf 93}, 160401 (2004).


\bibitem{Yuzbashyan2005}
E.A. Yuzbashyan, B.L. Altshuler, V.B. Kuznetsov, and V.Z. Enolskii
``Nonequilibrium Cooper pairing in the nonadiabatic regime,''
Phys. Rev. B {\bf 72}, 220503 (2005),
arXiv:cond-mat/0505493.


\bibitem{Yuzbashyan2008}
E.A. Yuzbashyan,
``Normal and anomalous solitons in the theory of dynamical Cooper pairing,''
Phys. Rev. B {\bf 78}, 184507 (2008),
arXiv:0807.3181.


\bibitem{Gurarie2009}
V. Gurarie,
``Nonequilibrium dynamics of weakly and strongly paired superconductors,''
Phys. Rev. Lett. {\bf 103}, 075301 (2009),
arXiv:0905.4498.


\bibitem{Gurarie2015}
E.A. Yuzbashyan, M. Dzero, V. Gurarie, and M.S. Foster,
``Quantum quench phase diagrams of an s-wave BCS--BEC condensate,''
Phys. Rev. A {\bf 91}, 033628 (2015),
arXiv:1412.7165.


\bibitem{Matsunaga2013}
R. Matsunaga, Y.I. Hamada, K. Makise, Y. Uzawa, H. Terai, Z. Wang,
and R. Shimano,
``The Higgs amplitude mode in BCS superconductors Nb$_{1-x}$Ti$_{x}$N induced by terahertz pulse excitation,''
Phys. Rev. Lett. {\bf 111}, 057002 (2013), arXiv:1305.0381.


\bibitem{Matsunaga2014}
R. Matsunaga, N. Tsuji, H. Fujita, A. Sugioka, K. Makise, Y. Uzawa, H.
Terai,  Z. Wang, H. Aoki, and R. Shimano,
``Light-induced collective pseudospin precession resonating with Higgs mode in a superconductor,''
Science {\bf 345}, 1145 (2014).


\bibitem{KlinkhamerVolovik2008a}
F.R. Klinkhamer and G.E. Volovik,
``Self-tuning vacuum variable and cosmological constant,''
Phys. Rev. D \textbf{77}, 085015 (2008), arXiv:0711.3170.

\bibitem{KlinkhamerVolovik2008b}
F.R. Klinkhamer and G.E. Volovik,
``Dynamic vacuum variable and equilibrium approach in cosmology,''
Phys. Rev. D \textbf{78}, 063528 (2008), arXiv:0806.2805.

\bibitem{KlinkhamerVolovik2010}
F.R.~Klinkhamer and G.E.~Volovik,
``Towards a solution of the cosmological constant problem,''
JETP Lett.\  {\bf 91}, 259 (2010),
arXiv:0907.4887.  

\bibitem{KlinkhamerVolovik2016-JETPL}
F.R. Klinkhamer and G.E. Volovik,
``Brane realization of $q$-theory and the cosmological constant problem,''
JETP Lett. {\bf 103}, 627 (2016),
arXiv:1604.06060.



\bibitem{DuffNieuwenhuizen1980}
M.J. Duff and P. van Nieuwenhuizen,
``Quantum inequivalence of different field representations,''
Phys. Lett.  B {\bf 94}, 179 (1980).

\bibitem{Aurilia-etal1980}
A. Aurilia, H. Nicolai, and P.K. Townsend,
``Hidden constants: The theta parameter of QCD and the cosmological constant of $N=8$ supergravity,''
Nucl.\ Phys.\  B {\bf 176}, 509 (1980).

\bibitem{Hawking1984}
S.W. Hawking,
``The cosmological constant is probably zero,''
Phys. Lett.  B {\bf 134}, 403 (1984).

\bibitem{HenneauxTeitelboim1984}
M. Henneaux and C. Teitelboim,
``The cosmological constant as a canonical variable,''
Phys.\ Lett.\  B {\bf 143}, 415 (1984).

\bibitem{Duff1989}
M.J. Duff,
``The cosmological constant is possibly zero, but the proof is probably wrong,''
Phys.\ Lett.\  B {\bf 226}, 36 (1989).

\bibitem{DuncanJensen1989}
M.J. Duncan and L.G. Jensen,
``Four-forms and the vanishing of the cosmological constant,''
Nucl.\ Phys.\  B {\bf 336}, 100 (1990).

\bibitem{BoussoPolchinski2000}
R. Bousso and J. Polchinski,
``Quantization of four-form fluxes and dynamical neutralization of the   cosmological constant,''
JHEP {\bf 0006}, 006 (2000), arXiv:hep-th/0004134.

\bibitem{Aurilia-etal2004}
A. Aurilia and E. Spallucci,
``Quantum fluctuations of a `constant' gauge field,''
Phys.\ Rev.\  D {\bf 69}, 105004 (2004), arXiv:hep-th/0402096.

\bibitem{Wu2008}
Z.C. Wu,
``The cosmological constant is probably zero, and a proof is possibly right,''
Phys. Lett.  B {\bf 659}, 891 (2008), arXiv:0709.3314.


\bibitem{ZeldovichStarobinsky1977}
Ya.B. Zel'dovich and A.A. Starobinsky,
``Rate of particle production in gravitational fields,''
JETP Lett. \textbf{26}, 252 (1977).


\bibitem{BirrellDavies1982}
N.D. Birrell and P.C.W. Davies,
\emph{Quantum Fields in Curved Space}
(Cambridge University Press, Cambridge, England, 1982).


\bibitem{Kofman1997}
L. Kofman, A. Linde, and A.A. Starobinsky,
``Towards the theory of reheating after inflation,''
Phys. Rev. D \textbf{56}, 3258 (1997).


\bibitem{DobadoMaroto1999}
A. Dobado and A.L. Maroto,
``Particle production from non-local gravitational effective action,''
Phys.\ Rev.\  D {\bf 60}, 104045 (1999), arXiv:gr-qc/9803076.


\bibitem{Klinkhamer2012}
F. R. Klinkhamer,
``On vacuum-energy decay from particle production,''
Mod. Phys. Lett. A \textbf{27}, 1250150 (2012),
arXiv:1205.7072.  


\bibitem{KlinkhamerVolovik2016-MPLA}
F.R. Klinkhamer and G.E. Volovik,
``Dynamic cancellation of a cosmological constant and approach to the Minkowski vacuum,''
Mod. Phys. Lett. A {\bf 31}, 1650160 (2016),
arXiv:1601.00601. 



\bibitem{GurovichStarobinsky1979}
V.Ts. Gurovich and A.A. Starobinsky,
``Quantum effects and regular cosmological models.''
Sov. Phys. JETP {\bf 50}, 844 (1979).

\bibitem{Starobinsky1981}
A.A. Starobinsky,
``Can the effective gravitational constant become negative?,''
Sov. Astron. Lett. {\bf 7}, 36 (1981).

\bibitem{Felice2012}
A. De Felice, K. Karwan, and P. Wongjun,
``Reheating in three-form inflation,''
Phys. Rev. D \textbf{86}, 103526 (2012),
arXiv:1209.5156.  


\bibitem{Akhmedov2014}
E.T. Akhmedov,
``Lecture notes on interacting quantum fields in de Sitter space,''
Int. J.  Mod. Phys. D {\bf 23}, 1430001 (2014),
arXiv:1309.2557.


\bibitem{SteinhardtTurok2002}
P.J. Steinhardt and N. Turok,
``A cyclic model of the universe,"
Science {\bf 296}, 1436 (2002), arXiv:hep-th/0111030.

\end{thebibliography}
\end{document}